\documentclass[sigconf]{acmart}
\usepackage{tikz}
\usetikzlibrary{shapes.geometric}
\usepackage{xcolor}
\usepackage{booktabs}  
\usepackage{listings}
\usepackage{url}
\usepackage{hyperref}
\hypersetup{
  colorlinks=false,    
  urlcolor=blue       
}

\newcommand{\redstar}{\tikz[baseline=-0.6ex]\node[star,star points=5,star point ratio=2,fill=red!75!black,minimum size=7pt,inner sep=0pt]{};}

\usepackage{algorithm}
\usepackage{algorithmicx}
\usepackage{algpseudocode}
\usepackage{amsmath} 

\usepackage{enumitem}
\setcopyright{none}
\renewcommand\footnotetextcopyrightpermission[1]{}

\begin{document}

\title{DBcover: A White-box SQL Test Generation Framework for Coverage Improvement}

\author{Yankai Rong}
\affiliation{%
  \institution{Renmin University of China}
  \city{Beijing}
  \country{China}}
\email{yankairong@ruc.edu.cn}

\author{Shuang Liu}
\authornote{Corresponding author.}
\affiliation{%
  \institution{Renmin University of China}
  \city{Beijing}
  \country{China}}
\email{Shuang.Liu@ruc.edu.cn}

\author{Jinhao Dong}
\affiliation{%
  \institution{Renmin University of China}
  \city{Beijing}
  \country{China}}
\email{jhdong@stu.pku.edu.cn}

\author{Qiang Yin}
\affiliation{%
  \institution{China Electronics Technology Kingbase (Beijing) Technologies Inc.}
  \city{Beijing}
  \country{China}}
\email{qyin@kingbase.com.cn}

\author{Wei Lu}
\affiliation{%
  \institution{Renmin University of China}
  \city{Beijing}
  \country{China}}
\email{lu-wei@ruc.edu.cn}

\author{Jianhua Wang}
\affiliation{%
  \institution{China Electronics Technology Kingbase (Beijing) Technologies Inc.}
  \city{Beijing}
  \country{China}}
\email{jhwang@kingbase.com.cn}

\author{Xiaoyong Du}
\affiliation{%
  \institution{Renmin University of China}
  \city{Beijing}
  \country{China}}
\email{duyong@ruc.edu.cn}

\begin{abstract}

Relational Database Management Systems (RDBMSs) form the backbone of modern data-intensive applications, making their reliability and robustness of paramount importance. However, due to the large code base and immense code complexity, achieving high coverage in RDBMS testing remains a formidable challenge. Traditional fuzzing approaches rely on random SQL generation, which fails to capture the intricate correspondence between SQL and paths (SQL-to-path correspondence) within the codebase, leading to low code coverage. Symbolic execution–based methods can, in principle, address this issue but incur prohibitive computational costs and suffer from scalability bottlenecks on large systems.

We propose DBcover, an LLM-driven database test generation framework that performs white-box, code-aware SQL test generation through contextual reasoning. DBcover integrates lightweight dynamic analysis to extract SQL-to-path correspondence and call graphs as global context, and constructs source-level dependency information around target functions as local context. Both types of contextual information are organized within a unified knowledge graph, enabling efficient retrieval, reasoning, and reuse across iterative test generation.
Built upon this contextual foundation, DBcover adopts a two-phase test generation strategy. In the seed selection phase, it identifies a semantically relevant existing test case by analyzing SQL-to-path relationships within the knowledge graph. In the query generation phase, the LLM, equipped with both global and local context, employs a fine-grained reasoning pipeline to synthesize new SQL test cases that accurately trigger previously uncovered code regions.  Experimental results show that DBcover substantially outperforms existing fuzzers, achieving $80.1\%$ and $82.3\%$ coverage on PostgreSQL and MySQL. Experiments conducted on an enterprise RDBMS KingbaseES demonstrate the effectiveness and practical applicability of our approach on closed-source RDBMSs.

\end{abstract}

\keywords{DBMS, Large Language Model, Testcase Generation}

\maketitle
\pagestyle{plain}

\section{Introduction}
\label{sec:introduction}
Relational Database Management Systems (RDBMSs) serve as the core infrastructure of modern data-intensive applications, where their stability directly determines the reliability of upper-layer services. As a key metric for assessing testing adequacy, code coverage has been widely adopted to evaluate the completeness of system testing \cite{wang2019sensitive,zhong2020squirrel,wang2021industry}, exerting a direct impact on the quality assessment and release decisions of industrial RDBMSs. Given the pervasive deployment of RDBMSs in critical systems, improving code coverage of RDBMSs has become an urgent requirement for ensuring system quality and robustness.

\begin{figure*}[t]
  \centering
  \includegraphics[width=0.9\linewidth]{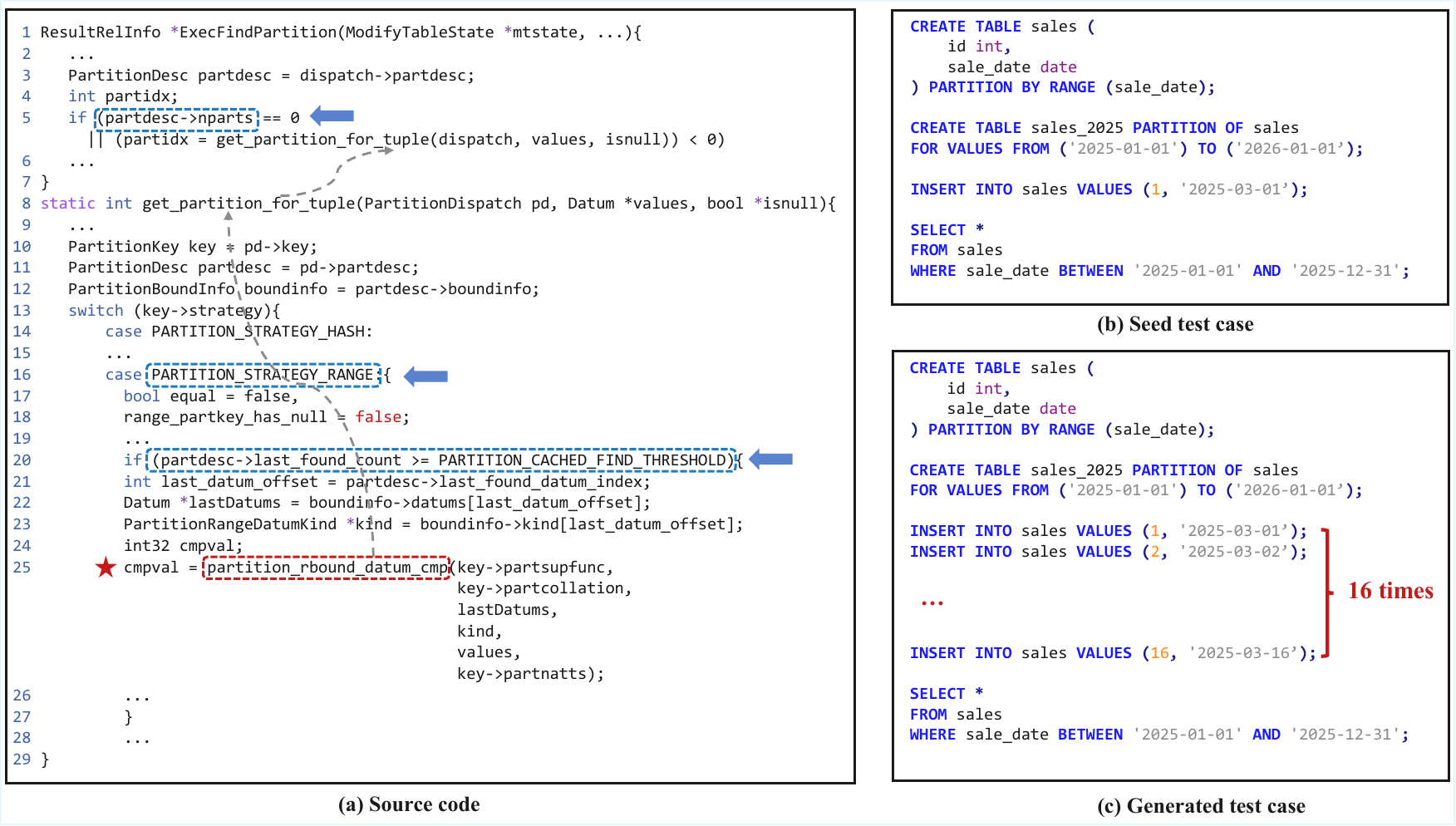}
  \caption{A Motivation example.}
  \label{fig:example_sql}
\end{figure*}

In recent years, a variety of automated SQL test case generation techniques have been proposed~\cite{zhong2020squirrel,sqlright,ba2023testing,dynsql}, aiming to produce both syntactically and semantically valid queries. Some approaches operate at the intermediate representation (IR) level to preserve semantic correctness during mutation~\cite{zhong2020squirrel,sqlright}, while others employ stateful fuzzing to maintain consistent database states throughout execution~\cite{dynsql}. Another line of work emphasizes fine-grained coverage collection~\cite{wang2021industry}, improving the applicability of testing frameworks in industrial settings.
Despite these advances, DBMS-oriented test generation techniques still face \textbf{a fundamental challenge that constrains their coverage improvement: the absence of source code level context information, i.e., an explicit mapping between SQL test cases and their corresponding execution paths (SQL-to-path correspondence)}. Existing gray-box methods~\cite{zhong2020squirrel,wang2021industry} provide limited coverage feedback but fail to capture the underlying code-level insights explaining why certain queries expand coverage, nor can they effectively guide the generation of new test inputs to expose untested paths. In principle, white-box approaches such as symbolic execution~\cite{symbolic} can establish such SQL-to-path correspondences. However, the immense complexity and scale of DBMS codebases lead to severe path explosion, rendering exhaustive symbolic analysis impractical.

\sloppy

Recent advances in Large Language Models (LLMs) have opened new possibilities for automated test generation. Several LLM-based approaches \cite{whitefox,kernelgpt,oliinyk2024fuzzing} leverage the models’ strong code comprehension capabilities by embedding the source code of target functions into prompts to generate function-level test cases. These methods typically focus on isolated modules or components of open source software systems—such as compiler optimization passes~\cite{whitefox}—where the context is well-bounded and can be explicitly provided to the model. Moreover, LLMs with strong capability, such as GPT-4 \cite{gpt4}, are adopted in those approaches.
Testing an RDBMS introduces fundamentally different challenges. These systems comprise millions of lines of code distributed across tightly coupled modules—such as query optimization, transaction management, and storage engines—with complex runtime dependencies and dynamic state interactions. Producing a test case that merely exercises a function’s general behavior is easy; generating one that precisely triggers a specific execution path or deep code location is far more difficult. Compounding this difficulty, industrial testing environments often enforce strict security policies (e.g., closed-source constraints) and limited computational resources (e.g., few available GPUs), which restrict the use of very large models (e.g., >70B parameters). Under these practical constraints, the central challenge becomes: how to effectively guide LLMs to generate test cases that increase code coverage in large-scale, highly entangled RDBMS codebases despite limited model capacity.

\textbf{Efficiently collecting the necessary contextual information from the RDBMS to guide SQL test case generation is a central challenge.} This observation motivates the design of DBcover, an LLM-driven white-box SQL test generation framework that employs a two-layer context collection strategy. DBcover explicitly models both the global and local contexts of the target code and unifies them within a comprehensive knowledge graph.

DBcover operates in two phases. In the first phase, it collects the global context, which includes the call graph and the SQL-to-path correspondence. The latter is obtained by executing existing SQL test cases—such as those from regression suites—and recording their detailed execution traces. These traces establish a fine-grained mapping between SQL inputs and the internal code paths they exercise. Using this mapping, DBcover computes the distance between uncovered code regions and existing execution paths in the call graph, and then identifies the closest covered path along with its associated test case to serve as the seed for subsequent test generation.
In the second phase, DBcover collects local context, including the source code of the target function and its dependency information, to guide the LLM in generating new test cases. To enable the model to better focus on problem-solving and mitigate the impact of model hallucinations, DBcover designs a task-specific pipeline, decomposing the overall task into five fine-grained subtasks, each supported by specialized prompt templates.
Through this two-phase process, DBcover effectively transforms the challenge of contextualizing the RDBMS system into a hierarchical reasoning problem. By integrating the global SQL-to-path correspondence context and the local source code context, it enables LLMs to generate semantically valid, coverage-oriented SQL test cases with high accuracy.
The main contributions of our study are as follows:

\begin{itemize}[left=0.7em]
\item We propose DBcover, an LLM-driven white-box test case generation framework for RDBMSs that integrates lightweight program analysis with LLM reasoning to achieve effective code coverage improvement.
\item 	We propose a two-layer context collection strategy for efficiently collecting the necessary contexts. (1) The global context layer captures SQL-to-path correspondence from existing test cases via dynamic execution analysis, providing a mapping between SQL statements and their triggered internal code paths. (2) The local context layer extracts detailed target code and dependency information to guide LLM reasoning. Both layers are organized within a unified knowledge graph, enabling efficient context retrieval, reasoning, and reuse across iterative test generation.
\item We propose a two-phase test generation strategy built upon the collected contextual knowledge. (1) In the seed selection phase, DBcover queries the knowledge graph to identify the seed test case whose execution path is closest to the target function in the call graph. (2) In the query generation phase, the LLM, equipped with both global and local context retrieved from the knowledge graph, employs a fine-grained reasoning pipeline to synthesize new SQL test cases designed to traverse previously uncovered code regions precisely.
\item We conduct extensive experiments on PostgreSQL, MySQL, and a real-world industrial closed-source DBMS (KingbaseES). Results show that DBcover significantly outperforms state-of-the-art SQL test generation methods in line coverage.
\end{itemize}

\section{Motivation Example}
\label{sec:motivatingexmp}

Fig.~\ref{fig:example_sql} illustrates a motivation example, including the source code of uncovered code lines (Fig.~\ref{fig:example_sql}(a)), the seed test case (Fig.~\ref{fig:example_sql}(b)) we selected from the regression tests, and the test case generated by our approach (Fig.~\ref{fig:example_sql}(c)), which can cover the target line, i.e., the function \texttt{partition\_rbound\_datum\_cmp} (highlighted with a \redstar~in line 25 of Fig.~\ref{fig:example_sql}(a)). It is a cache optimization strategy that can only be triggered when the system repeatedly accesses the same partition over a given threshold, i.e., the condition \texttt{partdesc->last\_found\_count} >= \texttt{PARTITION\_CACHED\_FIND\_THRESHOLD} in line 20 evaluates to true.
To reach the condition in line 25, the table partition strategy must be partition-by-range (\texttt{PARTITION\_STRATEGY\_RANGE} evaluates to true in line 16).
If we trace a bit further beyond the current function (\texttt{get\_partition\_for\_tuple}), the condition \texttt{partdesc->nparts == 0} in line 5 must evaluate to false (indicating that at least one valid partition descriptor exists), otherwise the optimization strategy in the C compiler will skip evaluating the second part of the \texttt{||} operator and fail to trigger the function \texttt{get\_partition\_for\_tuple}.
If we trace further back along the call chain, more complex path conditions will be retrieved. This is how the symbolic execution approach works.

Fig.~\ref{fig:example_sql}(b) shows a seed test case that exercises line 20 but fails to reach line 25. Inspection of the SQL statements reveals that the seed already satisfies the code-level semantics we derived from the source: the table is range-partitioned and contains a valid partition descriptor. Nonetheless, producing this seed via existing automatic generators is non-trivial: the seed comprises multiple SQL statements with intricate inter-statement value dependencies and precise semantic constraints, which typical syntax-oriented or naive fuzzing approaches struggle to satisfy.
Covering the deeper target at line 25 is substantially harder. Fig.~\ref{fig:example_sql}(c) shows the test case generated by DBcover, which can hit line 25. Achieving this coverage requires not only crafting semantically correct statements but also executing repeated insertions against the same partitioned table across multiple runs so that runtime counters/state (e.g., cached-find counts or access-threshold conditions) evolve to the values that trigger the desired path. Existing generation methods do not capture such internal program states and thus fail to systematically synthesize inputs that reach these kinds of guarded, stateful code regions.

DBcover generates the test case shown in Fig.~\ref{fig:example_sql}(c) through a two-phase process. In the first phase, DBcover analyzes existing regression test cases to extract the SQL-to-path correspondence by recording their execution traces and constructing a call graph centered on the target function. These structural and dynamic insights are then organized into a knowledge graph, which captures the relationships among SQL statements, execution paths, and invoked code entities—providing a structured and interpretable global context for test generation.
Given a specific uncovered line (e.g., line 25), DBcover traverses the call graph to locate the execution path with the smallest distance to the function \texttt{get\_partition\_for\_tuple}, which contains the target code region. The test case associated with this nearest execution path is then selected as the seed test case. This seed provides both the semantic and structural foundation for subsequent generation, guiding the LLM in synthesizing new SQL statements that are more likely to trigger the uncovered target line.

\begin{figure*}[t]
  \centering
  \includegraphics[width=1.0\linewidth]{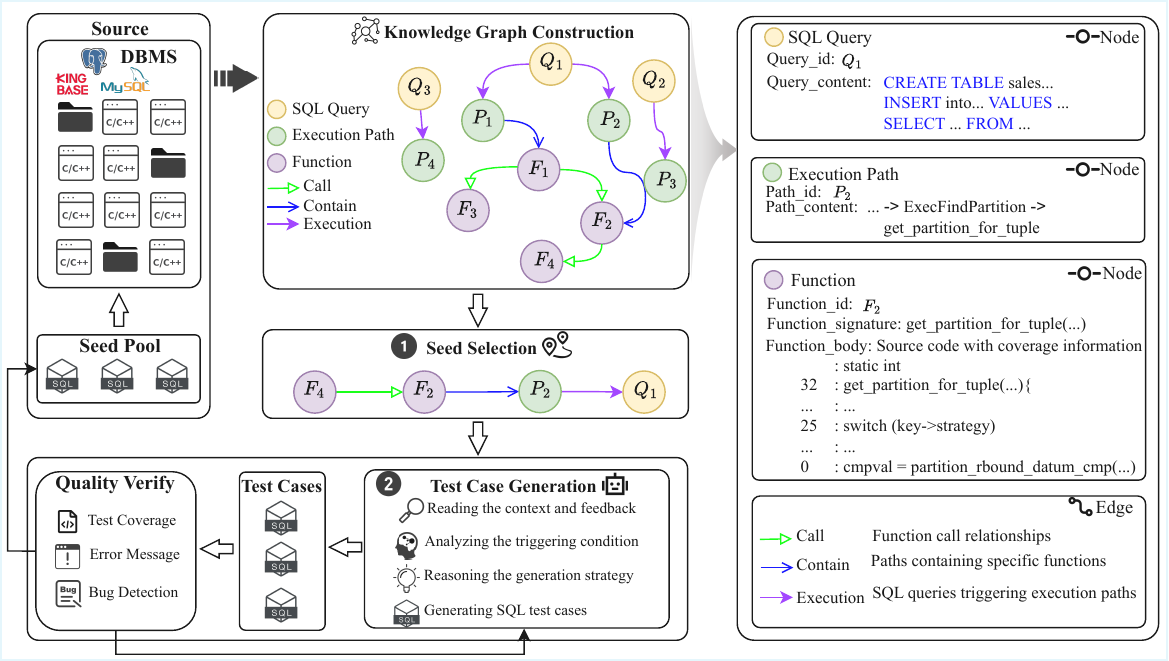}
  \caption{Overview of the DBcover framework.}
  \label{fig:overview}
\end{figure*}
\vspace{10pt}
In the second phase, DBcover leverages the code comprehension and reasoning capabilities of large language models (LLMs) to synthesize a new test case that can trigger the target uncovered line. This phase is guided by the global context extracted earlier, including the seed test case and its execution path, together with the local context, i.e., the source code of the target function. To enhance reasoning quality and controllability, DBcover adopts a  fine-grained reasoning pipeline, decomposing the overall task into five fine-grained subtasks, each with targeted prompt instructions.

\begin{enumerate}[left=0.7em]
    \item The LLM is first provided with the seed SQL test case (Fig.~\ref{fig:example_sql}(b)) and its associated execution path, which includes the call sequence from \texttt{get\_partition\_for\_tuple} to \texttt{partition\_rbound\_datum\_cmp}. From this global context, the model analyzes the high-level DBMS operations in the seed—specifically, the creation of a RANGE-partitioned table and a single tuple insertion—and infers the functional roles of routines along the call path from their names and invocation order.

    \item Next, the LLM examines the source code of the target function \texttt{partition\_rbound\_datum\_cmp} and its immediate caller \texttt{get\_partition\_for\_tuple}, including function signatures and function body with coverage information.

    \item Integrating global and local insights, the LLM deduces the precise runtime conditions required to reach the uncovered code: (i) the table must use range partitioning, and (ii) the same partition must be accessed at least 16 times consecutively, as \texttt{PARTITION\_CACHED\_FIND\_THRESHOLD} is defined as 16.

    \item  Building on the insights derived from the preceding analysis, the LLM identifies that the seed satisfies the first condition but fails the second—it performs only one insertion, so \texttt{last\_found\_count} never reaches the threshold. Consequently, the cache-optimized path invoking \texttt{partition\_rbound\_datum\_cmp} remains unexecuted. To close this semantic gap, the LLM formulates a minimal, targeted strategy: augment the seed test with additional \texttt{INSERT} statements that map to the exact same RANGE partition as the original tuple. Specifically, 15 more insertions are needed to bring the total access count to 16, thereby satisfying the caching precondition and enabling execution of the target branch.

    \item Finally, the fifth subtask focuses on test case synthesis. Guided by the generation strategy, the LLM generates the new SQL test case shown in Fig.\ref{fig:example_sql}(c) based on the seed test case in Fig.\ref{fig:example_sql}(b). This SQL extends the seed test case by performing 15 additional insertions into the same RANGE partition, ensuring that the consecutive-hit condition is met. When executed, the generated test successfully triggers the cache-optimized branch, invokes \texttt{partition\_rbound\_datum\_cmp}, and covers the previously unreachable line 25.

\end{enumerate}

\section{Methodology}
\label{sec:method}
\subsection{Overview}
In this section, we present the design of DBcover. The core idea is to efficiently gather the contextual information needed from the RDBMS codebase so that an LLM can generate test cases that effectively improve code coverage. DBcover operates in two phases. First, it leverages existing regression test cases to extract global SQL-to-path context, enabling the identification of a seed test case whose execution reaches code regions closest to the uncovered target. Then, equipped with this seed, DBcover uses the LLM to interpret the local context—including the target function and its dependencies—and mutate the seed to synthesize new SQL inputs that more precisely exercise the uncovered code.

The workflow of DBcover is illustrated in Fig. \ref{fig:overview}. DBcover consists of three major components: knowledge graph construction, seed selection, and test case generation. DBcover first collects both global and local context information and unifies them into a knowledge graph. To minimize context collection cost, DBcover leverages the existing regression test suite of the target RDBMS as the initial seed pool. Each test case is executed once to obtain its execution path, and the resulting SQL-to-path traces are stored in the knowledge graph for reuse. Whenever new test cases, those that successfully improve coverage, are added to the seed pool, their execution paths are incrementally added to keep the knowledge graph up to date. In addition, DBcover automatically extracts the target function and its dependency information as local context and inserts them into the graph.
Built upon this knowledge graph, DBcover performs a two-phase test generation workflow. The seed selection module retrieves the seed whose execution path is closest to the target code region. Then, using this seed together with the global and local context, DBcover prompts the LLM with a fine-grained reasoning pipeline to synthesize new SQL inputs intended to cover the target code.

\subsection{Knowledge Graph Construction}
\label{sec:Construction_knowledge}
To effectively contextualize the RDBMS code base, we propose a two-layer context extraction strategy and organize the extracted context into a knowledge graph. Before diving into the context collection details, we first formally define the knowledge graph we used in DBcover.

\begin{definition}[Code Context Knowledge Graph]
We define the code context knowledge graph as $\mathcal{G} = (V, E, \mathrm{Prop}_V, \mathrm{Prop}_E, f_V, f_E)$, where $V$ is the set of nodes, $E$ is the set of edges.  $\mathrm{Prop}_V$ is the set of properties associated with the nodes and $\mathrm{Prop}_E$ is the set of properties associated with the edges. $f_V: V \to \mathrm{Prop}_V$ maps each node to its set of properties. $f_E: E \to \mathrm{Prop}_E$ maps each edge to its set of properties.
\end{definition}
 Nodes in a knowledge graph consist of three types: $\mathcal{Q}$: SQL queries, $\mathcal{P}$: execution paths, and $\mathcal{F}$: functions.  For SQL query nodes, the properties include \texttt{Query\_id} and \texttt{Query\_content}, for execution path nodes, the properties include \texttt{Path\_id} and \texttt{Path\_content}, and for function nodes, the properties include \texttt{Function\_id}, \texttt{Function\_signature}, and \texttt{Function\_body}.
 Edges captures three types of relationships between the nodes, including $\text{Execution}$: from a $\mathcal{Q}$ node to a $\mathcal{P}$ node, meaning a SQL query exercises an execution path;  $\text{Contain}$: from a $\mathcal{P}$ node to a $\mathcal{F}$ node, meaning an execution path contains a function; and $\text{Call}$: from a $\mathcal{F}$ node to a $\mathcal{F}$ node, meaning one function calls another. For $\text{Execution}$ edges, the properties include \texttt{Execution\_id} and \texttt{Execution\_condition}, describing the condition under which a SQL query triggers an execution path. For $\text{Contain}$ edges, the properties include \texttt{Contain\_id} and \texttt{Contain\_relationship}, describing the relationship where an execution path contains a function. For $\text{Call}$ edges, the properties include \texttt{Call\_id} and \texttt{Call\_relation}, describing the relationship where one function calls another.

 Figure~\ref{fig:overview} illustrates the knowledge graph constructed by DBcover. The graph consists of three types of nodes, i.e., SQL queries, execution paths, and functions, and three types of directed edges: \text{Execution} from a SQL query to an execution path, \text{Contain} from an execution path to a function, and \text{Call} from one function to another.
 Each node in the graph is associated with a set of structured properties. For instance, the SQL query  $Q_1$ includes properties such as \texttt{Query\_id}: ``$Q_1$'' and \texttt{Query\_content}: ``CREATE TABLE t1 (a INT) PARTITION BY RANGE (a); ...'', representing a specific test case.
 The execution path node $P_2$ holds properties like \texttt{Path\_id}: ``$P_2$'' and \texttt{Path\_content}: ``ExecFindPartition -> get\_partition\_for\_tuple'.
 The function node  $F_2$ stores properties such as \texttt{Function\_id}: ``$F_2$'', \texttt{Function\_signature}: ``partition\_rbound\_datum\_cmp'' and the parameters, and \texttt{Function\_body}, where \texttt{Function\_body} includes the source code snippet annotated with coverage information.

 The edges in the knowledge graph represent the relationships between these nodes. $\text{Execution}$ edges link SQL queries to their corresponding execution paths. For example, the $\text{Execution}$ edge from $Q_1$ to $P_2$ reflects the pathability from a query to its runtime behavior. $\text{Contain}$ edges connect execution paths to the specific functions they exercise, such as the edge from $P_2$ to $F_2$, which grounds dynamic observations in static code elements. $\text{Call}$ edges represent internal function call dependencies, such as the edge from $F_1$ to $F_2$, revealing the internal structure of the DBMS.

The code context knowledge graph is constructed progressively during our test generation process. We categorize the code context into global context and local context, based on the scope of the context information. The global context contains information that spreads across a large portion of the project, for instance, the execution path of a given test case, the call graph, etc. The local context includes information that is local to our target function or code lines, for instance, the source code of the target function, its current coverage, etc.

\noindent\textbf{Global Context Extraction.}
The global context, including the SQL-to-path correspondence and the call graph, is extracted at the beginning, when we analyze the DBMS source code and the project’s directory structure. It can also be extracted incrementally during the test generation process. When new test cases are added to the seed pool, this process is automatically triggered.
This process begins by systematically extracting the source code and organizing it into logical units based on the relationships between function calls.
We will execute all the regression test cases in the seed pool to collect SQL-to-path correspondence information.
During the execution of SQL test cases, their execution paths are tracked in real time, capturing the dynamic interactions between the queries and the underlying code. These execution paths are represented as execution path nodes in $V$ and are connected to the relevant SQL query nodes through $\text{Execution}$ edges.
The execution paths, along with the function call graph, are integrated into the knowledge graph, linking SQL queries to their corresponding execution paths and functions. This approach ensures that we capture the broad execution context of the DBMS, providing a comprehensive view of system behavior.

\noindent\textbf{Local Context Extraction.}
The local context, in contrast, focuses on a specific target function and its source code augmented with coverage information. Formally, let $F_i \in V$ denote the function node corresponding to the target function in the knowledge graph $\mathcal{G}$.
Its associated properties, given by $f_V(F_i)$, include \texttt{Function\_id}, \texttt{Function\_signature}, and \texttt{Function\_body}, where \texttt{Function\_body} embeds line-level execution counts derived from prior test runs. To obtain this coverage-enriched representation, DBcover instruments the target DBMS to collect line-level execution counts during test runs and associates them with the corresponding source code of $F_i$. This representation precisely captures which lines within $F_i$ remain unexecuted or under-tested. By grounding test generation in this coverage-enriched local context, DBcover can systematically target unexplored code in the function.

This knowledge graph effectively expresses the relationship between SQL queries, execution paths, and the underlying code functions. It integrates dynamic execution information with static code structure, providing strong semantic support for subsequent test case generation and path analysis. The knowledge graph serves as a rich representation, allowing for precise and targeted test case generation based on the execution context and coverage data.

\begin{algorithm}[t]
\caption{Seed Selection}
\label{alg:seed-selection}
\begin{algorithmic}[1]
\State \textbf{Input:} Knowledge graph $\mathcal{G}$, target function $F_{\text{target}}$, maximum depth $D$
\State \textbf{Output:} Seed SQL query or $\emptyset$
\For{$d = 0$ \textbf{to} $D$}
    \For{each function $f$ with $\mathrm{dist}(F_{\text{target}}, f) = d$}
        \If{$\exists\, P \in \mathcal{P} : f \in P$}
            \State \textbf{return} SQL query of $P$
        \EndIf
    \EndFor
\EndFor
\State \textbf{return} $\emptyset$
\end{algorithmic}
\end{algorithm}

\subsection{Seed Selection}
In coverage-guided testing of complex systems like RDBMSs, blindly generating SQL queries from scratch is often ineffective: the probability of randomly triggering a deep, target-specific function (e.g., a query optimizer routine) is extremely low due to intricate parsing, validation, and dispatch logic. To address this, DBcover leverages the knowledge graph $\mathcal{G}$ to perform \textit{informed seed selection}—identifying existing test cases that already execute code close to the target function $F_{\text{target}}$.

The key insight is that functions reachable via a few static call edges from the target function ($F_{\text{target}}$) are more likely to share execution context (e.g., data structures, transaction state, or query plan stages). Thus, small, semantics-preserving mutations to a seed that reaches such a neighboring function can often bridge the remaining gap to $F_{\text{target}}$. We quantify this notion of proximity using the shortest-path distance over \textsc{Call} edges in the knowledge graph  $\mathcal{G}$. Formally, we use $\mathrm{dist}(f, F_{\text{target}})$ to denote the minimum number of directed \textsc{Call} edges from function $f$ to the target function $F_{\text{target}}$ in the knowledge graph.

As shown in Algorithm~\ref{alg:seed-selection}, DBcover searches for the closest function $f$ to the target function $F_{target}$ (by $\mathrm{dist (F_{target}, f)}$), where $f$ is on the execution path of existing test cases recorded in the knowledge graph. Specifically, it iterates over distances $d = 0, 1, \dots, D$, where $D$ is a configurable bound that prevents excessive search in deeply nested call chains. For each $f$ at distance $d$, it checks whether any execution path $P \in \mathcal{P}$ contains $f$ (i.e., the dynamic path of some past test exercised $f$). If such a path exists, the algorithm returns its associated SQL query as the seed.

Figure~\ref{fig:overview} illustrates this process: given $F_4$ as the target function $F_{\text{target}}$, DBcover identifies a function $F_2$ that is close to it through the \textsc{Call} edge in the knowledge graph ($F_2 \to F_4$). It then finds an execution path $P_2$ that contains $F_2$ through the \textsc{Contain} edge in the knowledge graph ($P_2 \to F_2$), indicating that $F_2$ was actually executed during a previous test. This path is linked back to the original SQL query $Q_1$ via the \textsc{Execution} edge. Since $Q_1$ has already triggered $F_2$, it serves as an ideal seed for mutating toward $F_{\text{target}}$.

\begin{figure}[t]
  \centering
  \includegraphics[width=1.0\linewidth]{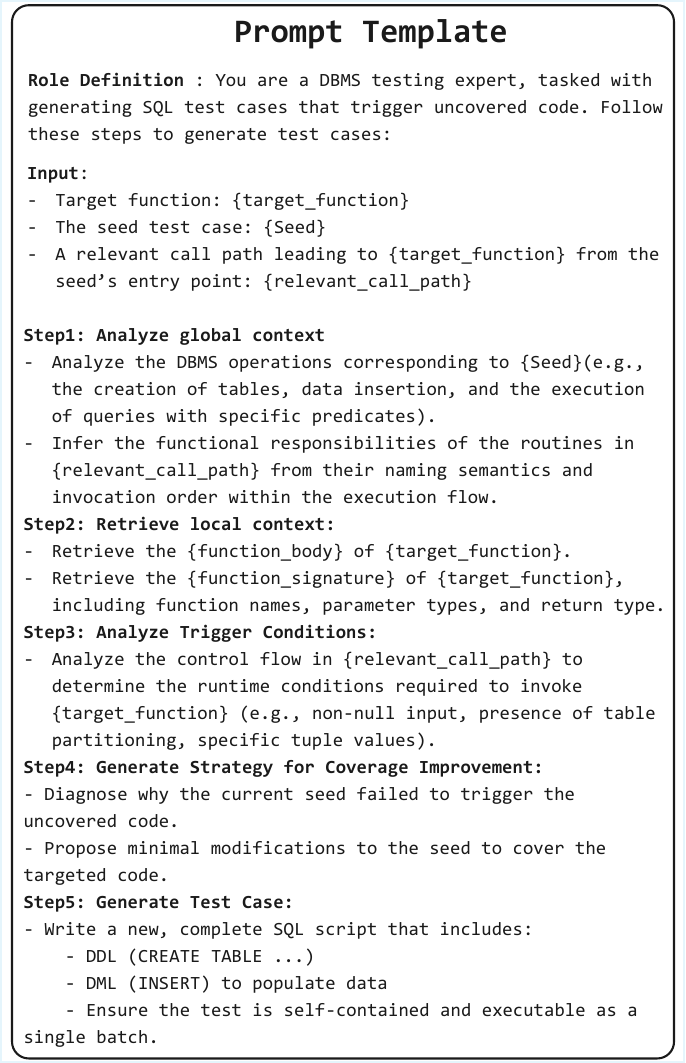}
  \caption{Fine-grained reasoning prompt template.}
  \label{fig:prompt}
\end{figure}
\vspace{10pt}

\subsection{Test Cases Generation}
The Test Case Generation component of DBcover leverages the reasoning capabilities of LLMs to synthesize SQL test cases that exercise previously uncovered RDBMS code regions. Instead of presenting the model with large volumes of undifferentiated source code, DBcover provides a curated contextual package extracted from the knowledge graph $\mathcal{G}$. To enhance reasoning accuracy and interpretability, particularly under the 32B-parameter LLM used in industrial deployment, DBcover organizes test generation into a fine-grained five-stage pipeline, with each stage grounded in information retrieved from $\mathcal{G}$. The model first interprets the global context. It then narrows its focus to the local context of the target function, including its body, signature, and uncovered regions. Anchored in this global-to-local progression, the LLM can accurately infer the specific runtime conditions required to reach the missing coverage and transform them into concrete SQL test cases. As shown in the prompt template in Fig.~\ref{fig:prompt}, the detailed steps are as follows:

\vspace{1mm}
\noindent\textbf{1. Analyze Global Context:}
The system first presents the LLM with the seed SQL script \texttt{\{Seed\}} and its associated execution path \texttt{\{relevant\_call\_path\}}. Guided by the prompt, the model (i) analyzes the DBMS operations performed by \texttt{\{Seed\}} (e.g., table creation, data insertion, and the execution of queries with specific predicates), and (ii) infers the functional responsibilities of the routines along \texttt{\{relevant\_call\_path\}} from their naming semantics and invocation order within the execution flow. This step reconstructs the global execution context in which the target function is embedded.

\vspace{1mm}
\noindent\textbf{2. Retrieve Local Context:}
After reconstructing the global execution environment, DBcover directs the LLM to examine the local code context of \texttt{\{target\_function\}}. The LLM retrieves the function’s body, signature, parameters, and uncovered regions. Based on this information, the model identifies the internal computation performed by the target routine, the meaning of its inputs and outputs, and how its control-flow guards relate to the surrounding execution path. This step anchors the reasoning in the precise code that needs to be triggered.

\vspace{1mm}
\noindent\textbf{3. Analyze Trigger Conditions:}
Using both the global context (from the seed and its execution path) and the local context of the target function, the LLM analyzes the control-flow dependencies leading to \texttt{\{target\_function\}}. It infers the runtime conditions necessary to reach the uncovered code—such as specific argument values, required table states, or repeated access patterns that activate caching or optimization paths. For example, when targeting \texttt{partition\_rbound\_datum\_cmp}, the model deduces that the cache-enabled branch becomes reachable only after repeated boundary comparisons in a test case meet or exceed the predefined threshold. This analysis reveals exactly why the seed fails to trigger the desired path.

\vspace{1mm}
\noindent\textbf{4. Generate Strategy for Coverage Improvement:}
The LLM then diagnoses the semantic gap between the inferred trigger conditions and the behavior of \texttt{\{Seed\}}. By comparing the required runtime state with what the seed actually constructs, the model proposes minimal and targeted modifications—such as generating additional inserts mapping to the same partition, issuing a more selective predicate, or restructuring the query to induce repeated lookups. This strategy outlines the concrete transformations needed for the seed to reach the previously uncovered region.

\vspace{1mm}
\noindent\textbf{5. Generate Test Case:}
Finally, DBcover instructs the LLM to instantiate the improvement strategy into a fully executable SQL script. The model synthesizes all necessary DDL statements, constructs the appropriate DML operations to populate data with the required structure and distributions, and formulates the triggering query that satisfies the inferred runtime conditions. The generated script is self-contained and ordered as a single batch, ensuring that all prerequisite states—such as table schemas, inserted tuples, and access patterns—are established within the test itself.

This staged, context-aware prompting enables systematic exploration of stateful, hard-to-reach DBMS logic—such as partition caching mechanisms—while remaining compatible with lightweight models and automated pipelines. After generation, synthesized test cases are executed under instrumentation (e.g., \texttt{gcov}). Feedback is analyzed along two axes: (1) whether the target branch is now covered, and (2) whether a crash or bug is exposed. Successful tests enrich the seed pool and update $\mathcal{G}$ with new \textit{Execution} edges and refined coverage. Failed attempts trigger iterative refinement: execution paths, runtime values, and error messages are incorporated into a revised prompt, enabling the LLM to diagnose and correct its reasoning in subsequent rounds. This closed-loop integration of graph-augmented context, structured reasoning, and dynamic feedback allows DBcover to progressively uncover deeper, state-dependent code paths in the DBMS.

\section{Evaluation}
In the evaluate section, we try to answer the following three research questions.

\noindent\textbf{RQ1:} How effective is DBcover in improving the test coverage of RDBMSs as compared to existing generation methods?

\noindent\textbf{RQ2:} How does the two-phase design of DBcover contribute to enhancing the effectiveness of DBcover?

\noindent\textbf{RQ3:} Is DBcover effective in closed-source industrial databases?

\subsection{Experimental Setup}
\subsubsection{Implementation}
To collect comprehensive data for test case generation, DBcover extracts function-level source information from the DBMS codebase using Doxygen \cite{doxygen}. To capture the function call relationships within the system, we compile the source code with LLVM, constructing a detailed call graph that represents the interactions between functions. For code coverage information, we use gcov and lcov, and we develop custom Python scripts to integrate coverage data from both source code and gcov files. SQL execution paths are traced by instrumenting the DBMS with the Pin tool \cite{pin}, enabling us to monitor the execution flow and capture the specific paths taken during query execution. In practice, Pin is employed exclusively offline for path collection, ensuring that generated tests execute natively in CI pipelines without incurring runtime overhead. All gathered data, including coverage metrics, function relationships, and execution traces, are stored in a Neo4j database \cite{neo4j}, providing an organized and queryable foundation for test case generation and evaluation.

\subsubsection{Environment}
For RQ1 and RQ2, the experiments were conducted on a server equipped with an Intel(R) Xeon(R) Gold 6454S Processor (128 cores) and 512GB of RAM, with 2 NVIDIA RTX 5880 Ada 48GB GPUs, running a 64-bit version of Ubuntu 22.04.5 LTS.
For RQ3, since it evaluates the performance of an industry closed-sourced RDBMS, we run the experiments  on their internal environments on a server equipped with an Intel(R) Core(TM) i9-14900K Processor (32 cores) and 64GB of RAM, with 2 NVIDIA GeForce RTX 4090 GPUs (each with 48GB of memory), running a 64-bit version of CentOS 7. Experiments for DBcover and its variant DBcover\textsubscript{nse} use the \texttt{qwen3:32b} large language model with a temperature of 0.7.

\subsubsection{Target RDBMSs}
To demonstrate the effectiveness of DBcover, we applied it to a variety of relational database management systems (DBMS), including the open-source systems MySQL (version 8.0.33) and PostgreSQL (version 17.0), as well as a proprietary industrial DBMS, KingbaseES \cite{kes}. The inclusion of both open-source and closed-source systems in our evaluation provides a comprehensive assessment of DBcover’s performance across different RDBMS implementations.

\subsubsection{Baselines:}
\begin{itemize}
    \item Base: The baseline approach refers to using the initial set of regression test cases as input, executing the test cases on the target RDBMS and measuring the collective coverage achieved after these executions.
    \item ShQveL \cite{zhong2025testing} enhances existing SQL test case generator (SQLancer++ \cite{sqlancerplus}) by leveraging LLM to synthesize SQL fragments. This is the first and only available LLM-assisted test generation method for RDBMS and thus is selected as a comparison method. In our implementation, we utilized Gemini-2.5-pro (with 175B parameters) as the LLM for the test case generation process.
    Additionally, we made modifications to certain parts of the code to adapt it to MySQL.
    \item SQUIRREL \cite{zhong2020squirrel}: is a representative test generation approach that conducts mutations on the intermediate representation (IR) representations, aiming to achieve high syntax and semantic correctness of the generated SQL query. SQUIRREL also supports the generation of multiple interdependent SQL statements. We supplemented squirrel's seed, thus maintaining the same initial seed as DBcover.
\end{itemize}

\subsubsection{Seed test cases}
Seed inputs were collected from the official GitHub repositories of the RDBMSs under evaluation. These repositories typically provide extensive regression test suites that cover a wide spectrum of query types, making them a strong foundation for initial test case selection. To ensure fairness, all compared methods are initialized with the same seed pool.

\begin{table}[t]
  \centering
  \caption{Line coverage achieved by the compared methods}
  \label{tab:coverage}
  \small 
  \setlength{\tabcolsep}{4pt} 
  \begin{tabular}{lccccc}
    \toprule
    DBMS & Base  & SQUIRREL & ShQveL & DBcover$_{\text{nse}}$ & DBcover \\
    \midrule
    PostgreSQL & 68.6\% & 69.0\% & 31.5\%   & 75.7\% & \textbf{80.1\%} \\
    MySQL      & 71.2\% & 71.4\% & 25.7\%   & 77.2\% & \textbf{82.3\%} \\
    \bottomrule
  \end{tabular}
\end{table}

\begin{table}[t]
\centering
\caption{Ablation study results (per module on PostgreSQL)}
\label{tab:coverage_summary}
\small
\setlength{\tabcolsep}{4pt}
\begin{tabular}{@{}lrrrr@{}}
\toprule
Project & Lines & Base & DBcover\textsubscript{nse} & DBcover  \\
\midrule
utils/adt          & 66211 & 81.60\% & 84.90\% & 88.00\% (+3.10\%, +6.40\%) \\
commands           & 32326 & 81.60\% & 83.40\% & 86.10\% (+2.70\%, +4.50\%) \\
nodes              & 24882 & 44.30\% & 44.70\% & 47.80\% (+3.10\%, +3.50\%) \\
parser             & 24206 & 87.30\% & 88.30\% & 90.00\% (+1.70\%, +2.70\%) \\
executor           & 23181 & 86.40\% & 86.50\% & 90.20\% (+3.70\%, +3.80\%) \\
catalog            & 13791 & 83.00\% & 85.90\% & 87.70\% (+1.80\%, +4.70\%) \\
access/transam     & 12118 & 52.80\% & 60.80\% & 64.60\% (+3.80\%, +11.80\%) \\
optimizer/plan     & 8434  & 91.10\% & 93.10\% & 96.20\% (+3.10\%, +5.10\%) \\
optimizer/util     & 7705  & 87.00\% & 87.90\% & 91.10\% (+3.20\%, +4.10\%) \\
access/heap        & 7241  & 67.10\% & 75.00\% & 83.90\% (+8.90\%, +16.80\%) \\
optimizer/path     & 7087  & 95.10\% & 95.20\% & 95.80\% (+0.60\%, +0.70\%) \\
access/nbtree      & 6310  & 86.90\% & 87.70\% & 88.00\% (+0.30\%, +1.10\%) \\
regex              & 5136  & 70.40\% & 73.00\% & 73.50\% (+0.50\%, +3.10\%) \\
tcop               & 4534  & 68.00\% & 70.50\% & 71.80\% (+1.30\%, +3.80\%) \\
postmaster         & 4299  & 52.60\% & 53.30\% & 53.90\% (+0.60\%, +1.30\%) \\
access/gin         & 4127  & 75.60\% & 77.20\% & 78.50\% (+1.30\%, +2.90\%) \\
utils/misc         & 3914  & 66.80\% & 71.10\% & 75.00\% (+3.90\%, +8.20\%) \\
tsearch            & 3643  & 88.70\% & 90.00\% & 90.20\% (+0.20\%, +1.50\%) \\
access/gist        & 3287  & 73.40\% & 75.80\% & 77.00\% (+1.20\%, +3.60\%) \\
utils/mmgr         & 3281  & 70.90\% & 72.10\% & 73.20\% (+1.10\%, +2.30\%) \\
access/brin        & 3198  & 70.20\% & 73.20\% & 84.30\% (+11.10\%, +14.10\%) \\
utils/activity     & 3127  & 66.30\% & 68.60\% & 72.30\% (+3.70\%, +6.00\%) \\
access/common      & 2849  & 78.10\% & 79.40\% & 80.40\% (+1.00\%, +2.30\%) \\
\midrule
\textbf{Average}   & 11952 & 75.01\% & 80.37\% & 83.08\% (+2.71\%, +8.07\%) \\
\bottomrule
\end{tabular}
\end{table}

\vspace{10pt}

\subsection{Comparison with Baselines}
\label{sec:comparison}

Table~\ref{tab:coverage} shows the results by each compared RDBMS test case generator in 24-hour time-bounded experiments. The column \textbf{Base} reports the coverage achieved by executing only the initial seed pool (i.e., without generating any new test cases).
We can observe that the initial seed pools already achieve a rather high line coverage, with \textbf{68.6\%} on PostgreSQL and \textbf{71.2\%} on MySQL.
DBcover achieves the highest coverage—\textbf{80.1\%} on PostgreSQL and \textbf{82.3\%} on MySQL—surpassing the seed baseline by \textbf{11.5\%} and \textbf{11.1\%}, respectively.
SQUIRREL improves coverage by merely \textbf{0.4\%} on PostgreSQL (68.6\% → 69.0\%) and \textbf{0.2\%} on MySQL.
ShQveL—a standalone LLM-augmented generator that does not leverage the seed corpus—achieves substantially lower coverage (\textbf{31.5\%} on PostgreSQL, \textbf{25.7\%} on MySQL).

We attribute the effectiveness of DBcover largely to its use of source-level context that explicitly links SQL inputs to their execution paths, which we refer to as \textit{SQL-to-path correspondence}. Without this mapping, test generators cannot determine which SQL constructs are needed to reach a given uncovered code region. Both SQUIRREL and shQveL lack such context information, and thus cannot effectively cover the target code region.
DBcover overcomes this barrier by explicitly collecting the SQL-to-path correspondence, providing LLM with this global context—and leveraging its native code comprehension ability—DBcover unlocks deep, previously unreachable code in production-grade DBMSs. Absent nearby execution paths, DBcover performs an upward call-graph search to locate the closest executed ancestor. Using this node as an anchor, the LLM infers the necessary state transitions to extend the execution path toward the deep target.

Through detailed analysis, we observe that most remaining uncovered code resides in regions that are fundamentally unreachable via valid SQL inputs during normal execution—such as crash-recovery routines, I/O error handlers, and system-management paths triggered only by operating-system–level events. Because DBcover (and all compared baselines) operates strictly through the user-facing SQL interface, these internal mechanisms lie outside the scope of what any SQL-based test generator can exercise. Beyond these architectural limits, a non-trivial fraction of uncovered code resides in SQL-reachable components but is guarded by complex path conditions requiring intricate runtime value computations. These conditions often depend on the results of deep internal function calls or comparisons against macro-defined system constants, and satisfying them typically demands precise control over multiple interdependent SQL parameters and system settings, which requires reasoning and numeric constraint solving beyond the current capabilities of small parameter LLMs. Together, these factors explain why overall coverage naturally plateaus around $80\%$.

\begin{figure*}[t]
  \centering
  \includegraphics[width=1.0\linewidth]{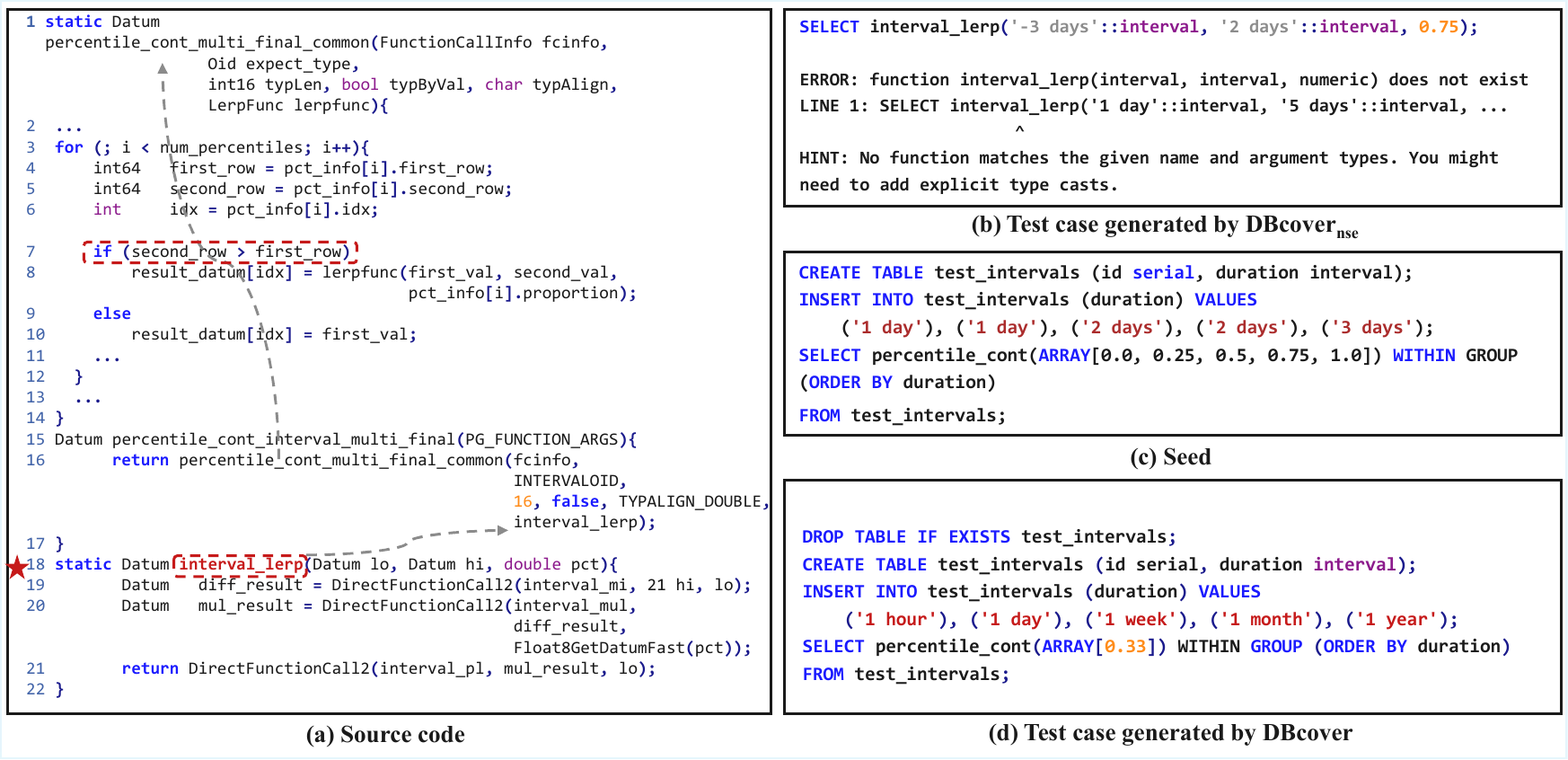}
  \caption{A Case Study.}
  \label{fig:casestudy}
\end{figure*}

\subsection{Ablation Study}
To evaluate the contribution of global context, we introduce \textbf{DBcover\textsubscript{nse}}, a variant of DBcover that removes the global context collection and seed selection steps. In \textbf{DBcover\textsubscript{nse}}, the LLM receives only the local context, i.e., the target function’s source code and line-level coverage information.

Table~\ref{tab:coverage} shows the evaluation results of running both settings of DBcover for 24 hours. DBcover\textsubscript{nse} achieves the coverage: $75.7\%$ on PostgreSQL and $77.2\%$ on MySQL, compared to DBcover’s $80.1\%$ and $82.3\%$, corresponding to drops of $4.4\%$ and $5.1\%$, respectively.
We further break down the coverage improvement on a per-module basis to assess the robustness of DBcover, i.e., whether the global context shows improvement consistently on different modules.
Table~\ref{tab:coverage_summary} summarizes the coverage improvement of top-23 largest PostgreSQL modules by source lines of code (SLOC), which together constitute the majority of the system’s core logic—including query execution (\texttt{executor}), planning (\texttt{optimizer/plan}, \texttt{optimizer/path}), and built-in data types (\texttt{utils/adt}). The last column of Table~\ref{tab:coverage_summary} shows three numbers, i.e., the coverage achieved by DBcover, its improvement over DBcover\textsubscript{nse}, and Base, respectively.

On average, DBcover achieves a 2.71\% coverage improvement over DBcover\textsubscript{nse} and 8.07\% over the Base configuration. These results demonstrate that both the seed selection mechanism and the fine-grained reasoning pipeline contribute significantly to generating test cases that effectively increase code coverage.
The largest coverage gain occurs in \texttt{access/brin} (+11.1\%, +14.1\%). The code in this module is guarded by strong runtime preconditions: a BRIN index must exist, the table must contain sufficient data to populate at least one range block, and a query referencing the indexed column must be issued. These conditions constitute a stateful dependency chain that can hardly be inferred from the target function’s source code. Therefore, the seed selection step shows great improvement.  What's more , the gap between DBcover\textsubscript{nse} and Base (average $+5.36\%$) reflects the benefit of local-code-aware prompting alone, while the additional $+2.71\%$ gain from DBcover highlights the unique contribution of our full prompt pipeline—specifically, its integration of global SQL-to-path context with structured, multi-step reasoning—to unlock stateful and semantically complex code paths that local analysis cannot reach.

A complementary benefit arises in modules governed by syntactic and semantic constraints rather than runtime state. For example, \texttt{utils/adt}—which implements PostgreSQL’s built-in data types such as \texttt{numeric}, \texttt{jsonb}, arrays, and ranges—improves from $84.9\%$ to $88.0\%$ (+3.1\%). Its functions are triggered only by specific SQL constructs, including type casts (e.g., \texttt{'123'::numeric}) and typed literals (e.g., \texttt{'[1,5]'::int4range}). While common cases are covered by the seed corpus, deeper branches—such as numeric overflow handling or recovery from malformed but syntactically valid JSON—require inputs that are both well-formed and corner cases. The seed provides a valid syntactic template, allowing the LLM to generate safe mutations (e.g., increasing precision or nesting depth) that remain parseable and type-consistent. Without this guidance, the LLM (DBcover\textsubscript{nse}) often produces ill-formed queries rejected by the parser before reaching the target code.

In contrast, several modules show minimal improvement. \texttt{optimizer/path} increases only from $95.2\%$ to $95.8\%$ (+0.6\%), \texttt{access/nbtree} from $87.7\%$ to $88.0\%$ (+0.3\%), and \texttt{tsearch} from $90.0\%$ to $90.2\%$ (+0.2\%). These components are already extensively exercised by the seed corpus, as their core logic is routinely triggered during standard query processing. The remaining uncovered code primarily consists of error-handling routines or assertions. Notably, the \texttt{postmaster} module also shows negligible gain (from $53.3\%$ to $53.9\%$, +0.6\%), but for a fundamentally different reason: it manages server lifecycle events such as process startup, client connection dispatch, and shutdown signaling. These behaviors are triggered by operating-system interactions or internal signals, not by SQL queries.

\subsection{Case Study}
Fig.~\ref{fig:casestudy} illustrates a case study on generating test cases that cover the target function \texttt{interval\_lerp} (line~19 of Fig.~\ref{fig:casestudy}(a)), including an erroneous test case generated without seed guidance (Fig.~\ref{fig:casestudy}(b)) and a correct test case synthesized by DBcover (Fig.~\ref{fig:casestudy}(d)) from the seed shown in Fig.~\ref{fig:casestudy}(c).

The function \texttt{interval\_lerp} is a static helper used internally by PostgreSQL’s percentile aggregation logic. When prompted with only the source code of \texttt{interval\_lerp}—without any seed test case—the LLM generates a syntactically plausible but semantically invalid query, as shown in Fig.~\ref{fig:casestudy}(b). The direct call \texttt{SELECT interval\_lerp(...)} fails when parsing because the function has no SQL interface, yielding the error: ``function interval\_lerp(interval, interval, numeric) does not exist''. The correct interface is ``\texttt{percentile\_cont}''.

In contrast, DBcover avoids this pitfall with a two-phase design. In the first stage, starting from the target function \texttt{interval\_lerp}, DBcover traverses the call graph upward and identifies \texttt{percentile\_cont\_multi\_final\_common} as the nearest function. Then the seed selection strategy searches for the corresponding seed, shown in Fig.~\ref{fig:casestudy}(c), in the knowledge graph. In the second stage, guided by a structured prompt, the LLM retrieves relevant context from the knowledge graph and analyzes the trigger condition for \texttt{interval\_lerp}. The function is invoked in line 8 by \texttt{lerpfunc}, which executes only when \texttt{second\_row > first\_row} (line~7 of Fig.~\ref{fig:casestudy}(a)) evaluates to true. Through code analysis, the LLM identifies that \texttt{first\_row} and \texttt{second\_row} are computed as $1 + \lfloor p \times (n-1) \rfloor$ and $1 + \lceil p \times (n-1) \rceil$ respectively, where $p$ is the percentile value and $n$ is the row count. The condition \texttt{second\_row > first\_row} holds true precisely when $p \times (n-1)$ is not an integer, indicating that the requested percentile falls between existing data points and requires interpolation. The LLM then diagnoses that the seed's percentiles—\texttt{[0.0, 0.25, 0.5, 0.75, 1.0]}—with $n=5$ result in $p \times (n-1) \in \{0, 1, 2, 3, 4\}$, all integers. Since $\lfloor k \rfloor = \lceil k \rceil = k$ for any integer $k$, \texttt{first\_row} equals \texttt{second\_row} for all percentiles, causing the condition in line~7 to evaluate to false. To satisfy the interpolation requirement, it generates a minimal, self-contained mutation by replacing the percentile array with \texttt{[0.33]}, where $0.33 \times 4 = 1.32$ (non-integer), ensuring $\lfloor 1.32 \rfloor = 1 < 2 = \lceil 1.32 \rceil$ and thus \texttt{second\_row} (3) > \texttt{first\_row} (2), successfully triggering \texttt{interval\_lerp}.
\vspace{-0.5em}

\subsection{Evaluation on Enterprise Closed-Source RDBMS}
We evaluate DBcover on \textbf{KingbaseES} \cite{kes}, a commercial relational DBMS widely used in critical enterprise systems. While KingbaseES is a closed-source product and its source code cannot be disclosed publicly, our evaluation is performed in an internal environment with a qwen3:32b model locally deployed where source-level instrumentation is permitted under a collaboration agreement.
Our testing targets two foundational subsystems that are central to the correctness and robustness of the database engine. The first is the \textit{data type processing module}, which implements built-in data types—such as numeric, string, and temporal types—as well as type coercion rules. The second is the \textit{metadata management module}, which governs all schema-related metadata.

Using DBcover’s test generation pipeline, we automatically synthesize SQL test cases that exercise complex behaviors in these subsystems—such as edge-case type conversions, nested expressions, and metadata-intensive operations. Starting from an initial coverage of 50.0\% in the data type processing module and 68.0\% in the metadata management module (achieved by existing internal test suites), DBcover significantly advances coverage to \textbf{80.0\%} and \textbf{81.4\%}, respectively. These results demonstrate that our approach can effectively explore deep, stateful logic in proprietary database engines. Due to licensing restrictions, internal implementation details cannot be disclosed; however, the reported coverage metrics reflect genuine execution paths within the production codebase.

\section{Related Work}
\subsection{RDBMS Test Case Generation Methods}
In traditional DBMS test case generation, fuzzing \cite{sqlright,dynsql,zhong2020squirrel,zhong2025testing,blazytko2019grimoire,wang2021industry,sqlancerplus} has received extensive attention due to its efficiency in automatically producing diverse inputs and triggering program behaviors. Existing approaches can be broadly grouped into two complementary directions, i.e., improving the syntactic and semantic validity of generated test cases, and refining coverage feedback to better reflect meaningful execution differences.

The first line of work focuses on generating well-formed SQL that adheres to DBMS grammar and semantics. Early tools like SQLsmith~\cite{sqlsmith2018}, SQLancer~\cite{sqlancer} and Apollo~\cite{apollo} generate single statements via syntax-driven traversal. Recent approaches produce multiple semantically related queries per test case, with generative methods synthesizing these groups from scratch. DynSQL~\cite{dynsql} dynamically queries the schema to guide abstract syntax tree instantiation into valid SQL.
SQUIRREL~\cite{zhong2020squirrel}, mutate seed queries, supporting syntax-preserving mutations. Griffin~\cite{fu2022griffin} recombines SQL fragments and repairs semantic inconsistencies using a metadata dependency graph. However, these methods treat the DBMS as a black/grey box and cannot reason about how SQL inputs map to internal execution paths, limiting their ability to improve code coverage.

Another line of work focuses on improving the accuracy of coverage feedback.
Traditional gray-box fuzzers rely on basic block or edge coverage~\cite{zhong2020squirrel,sqlright}, which conflates semantically distinct executions that follow the same control flow. Tools like Squirrel~\cite{zhong2020squirrel} and SQLRight~\cite{sqlright} extend AFL’s bitmap size (Squirrel enlarges the default 64 KB bitmap to 256 KB) to alleviate hash collisions and obtain finer-grained coverage feedback for SQL synthesis. Ratel~\cite{wang2021industry} addresses these issues by using static analysis to assign dedicated counters to basic blocks and branches, eliminating collisions, and by instrumenting all binaries at compile time for system-wide coverage in multi-process RDBMS. This line of work is orthogonal to ours. While this work does not generate queries, it provides valuable, precise coverage feedback that can enhance test-case generation guidance.

\vspace{-0.5em}

\subsection{LLM for Test Case Generation}
With the rapid progress of large language models (LLMs) in code understanding and generation, LLM-driven test case generation has surged recently \cite{whitefox,kernelgpt,yang2025autoverus,xu2025ckgfuzzer,zhang2025unlocking,coverup,mutap,advancing,zhong2025testing,shaohuali,xia2024fuzz4all}. Due to their strong semantic modeling abilities, LLMs can generate structured, semantically meaningful test inputs and have been integrated into various testing frameworks to enhance testing quality and automation across software systems. Two main approaches have emerged: using LLMs to augment existing test generators and directly synthesizing tests with LLMs.

In the augmentation paradigm, LLMs are integrated into existing test generators to enhance their generation process \cite{zhong2025testing,gai2025llama,luo2025enhancing}. For instance, ShQveL \cite{zhong2025testing} extracts SQL features through LLM interactions and incorporates them into existing generators, increasing behavioral coverage while maintaining efficiency. LLAMA~\cite{gai2025llama} uses an LLM to generate semantically valid seeds and refines them with multi-feedback optimization and hybrid fuzzing for smart contracts. In these approaches, LLMs assist with tasks like syntax repair, seed enrichment, and semantic expansion, compensating for the limitations of traditional generators. However, their effectiveness is still constrained by the base generator's ability to construct the execution context needed to trigger deep system logic, limiting improvements from LLM-based augmentation alone.

More recent research efforts \cite{whitebox,coverup,mutap,xia2024fuzz4all,cheng2025rug} increasingly treat large language models as primary test generators, producing complete and standalone test cases without relying on traditional mutation engines. For example, Whitefox \cite{whitefox} employs a multi-agent framework for testing compiler optimizations, progressively refining testing requirements to synthesize programs that activate specific optimization behaviors. CoverUp \cite{coverup} integrates coverage analysis, code context, and runtime feedback to iteratively generate new tests; however, its generation process does not leverage any seed traces obtained from real executions, leaving the model without guidance from concrete runtime semantics and resulting in exploratory generation under limited contextual constraints. MuTAP \cite{mutap} further combines mutation testing principles with prompt engineering, steering LLMs to produce Python unit tests and adding new assertions derived from surviving mutants to strengthen test quality. Despite promising results, existing LLM-based testing approaches typically depend on large-parameter models and are evaluated primarily on well-documented, open-source systems. Moreover, they tend to target isolated functionalities (e.g., compiler optimizations) rather than deeply integrated system software, whose complexity, dynamic interactions, and tight module interdependencies pose substantially greater challenges.

\section{Threats to Validity}

Threats to validity primarily concern the generalizability of our method across diverse database architectures and LLM backends. To ensure broad applicability, we conducted experiments on three representative RDBMSs, comprising PostgreSQL, MySQL, and KingbaseES. Among these, PostgreSQL and MySQL serve as the foundational open-source architectures for numerous derivative systems and are characterized by their immense codebase scale and logical complexity. KingbaseES complements this by representing mainstream commercial databases to ensure diversity in our evaluation. Methodologically, DBcover is inherently extensible across such diverse systems because its deterministic workflow is driven by generalizable program analysis and structured knowledge graphs rather than system-specific heuristics, ensuring a consistent and transferable testing process.

Regarding model dependence, our evaluation utilized a specific 32B-parameter model. The fact that even a relatively modest-sized code-capable LLM achieves significant gains when guided by precise and path-aware context indicates that the structured reasoning pipeline is the dominant factor in generating effective test cases, minimizing the reliance on extremely large model scales. This observation suggests that the choice of the underlying language model does not fundamentally compromise the validity of our findings. Moving forward, we plan to extend our evaluation to a broader range of LLMs, encompassing diverse parameter sizes and architectures, to further substantiate the framework’s robustness across varying model capabilities.

\section{Conclusion and Future Work}
 \label{sec:conclusion}
We propose DBcover, a LLM-driven database test generation framework that performs white-box, code-aware SQL test generation through contextual reasoning, aiming at effectively improving code coverage of RDBMSs. To tackle the challenge of efficiently contextualizing the RDBMS system, we propose a two-layer context collection strategy, which collects global context and local context related to the target function efficiently. We then propose a two-phase test generation approach, which relying on the global context to obtain a seed test case, and using local context to guide LLM for accurate test case generation. We conduct extensive experiments on PostgreSQL, MySQL, and a real-world industrial closed-source DBMS (KingbaseES). Results show that DBcover significantly outperforms state-of-the-art SQL test generation methods on code coverage.

Moreover, while our primary metric is code coverage, the path-aware SQL test cases generated by DBcover are inherently compatible with advanced bug-detection oracles, such as differential testing, metamorphic relations, and crash monitors, and provide a high-quality substrate for systematic bug discovery, which we plan to explore in future work.

\begin{acks}
This work is supported by the National Natural Science Foundation of China under Grant Nos. 62472429, 62441230, the Fundamental Research Funds for the Central Universities, and the Research Funds of Renmin University of China, and the A3 Foresight Program No. 62461146205. We gratefully acknowledge the support from the RUC-Kingbase Database Collaborative Innovation Joint Laboratory.
\end{acks}

\balance
\bibliographystyle{ACM-Reference-Format}
\bibliography{sample-base}

@String{Computer = "{IEEE} Computer" }

@inproceedings{wang2019sensitive,
  title={Be sensitive and collaborative: Analyzing impact of coverage metrics in greybox fuzzing},
  author={Wang, Jinghan and Duan, Yue and Song, Wei and Yin, Heng and Song, Chengyu},
  booktitle={22nd International Symposium on Research in Attacks, Intrusions and Defenses (RAID 2019)},
  pages={1--15},
  year={2019}
}

@inproceedings{zhong2020squirrel,
  title={Squirrel: Testing database management systems with language validity and coverage feedback},
  author={Zhong, Rui and Chen, Yongheng and Hu, Hong and Zhang, Hangfan and Lee, Wenke and Wu, Dinghao},
  booktitle={Proceedings of the 2020 ACM SIGSAC Conference on Computer and Communications Security},
  pages={955--970},
  year={2020}
}

@inproceedings{wang2021industry,
  title={Industry practice of coverage-guided enterprise-level DBMS fuzzing},
  author={Wang, Mingzhe and Wu, Zhiyong and Xu, Xinyi and Liang, Jie and Zhou, Chijin and Zhang, Huafeng and Jiang, Yu},
  booktitle={2021 IEEE/ACM 43rd International Conference on Software Engineering: Software Engineering in Practice (ICSE-SEIP)},
  pages={328--337},
  year={2021},
  organization={IEEE}
}

@inproceedings{sqlright,
  title={Detecting logical bugs of $\{$DBMS$\}$ with coverage-based guidance},
  author={Liang, Yu and Liu, Song and Hu, Hong},
  booktitle={31st USENIX Security Symposium (USENIX Security 22)},
  pages={4309--4326},
  year={2022}
}

@inproceedings{ba2023testing,
  title={Testing database engines via query plan guidance},
  author={Ba, Jinsheng and Rigger, Manuel},
  booktitle={2023 IEEE/ACM 45th International Conference on Software Engineering (ICSE)},
  pages={2060--2071},
  year={2023},
  organization={IEEE}
}

@article{whitefox,
  title={Whitefox: White-box compiler fuzzing empowered by large language models},
  author={Yang, Chenyuan and Deng, Yinlin and Lu, Runyu and Yao, Jiayi and Liu, Jiawei and Jabbarvand, Reyhaneh and Zhang, Lingming},
  journal={Proceedings of the ACM on Programming Languages},
  volume={8},
  number={OOPSLA2},
  pages={709--735},
  year={2024},
  publisher={ACM New York, NY, USA}
}

@inproceedings{kernelgpt,
  title={Kernelgpt: Enhanced kernel fuzzing via large language models},
  author={Yang, Chenyuan and Zhao, Zijie and Zhang, Lingming},
  booktitle={Proceedings of the 30th ACM International Conference on Architectural Support for Programming Languages and Operating Systems, Volume 2},
  pages={560--573},
  year={2025}
}

@inproceedings{oliinyk2024fuzzing,
  title={Fuzzing $\{$BusyBox$\}$: Leveraging $\{$LLM$\}$ and Crash Reuse for Embedded Bug Unearthing},
  author={Oliinyk, Yaroslav and Scott, Michael and Tsang, Ryan and Fang, Chongzhou and Homayoun, Houman and others},
  booktitle={33rd USENIX Security Symposium (USENIX Security 24)},
  pages={883--900},
  year={2024}
}

@inproceedings{dynsql,
  title={$\{$DynSQL$\}$: Stateful fuzzing for database management systems with complex and valid $\{$SQL$\}$ query generation},
  author={Jiang, Zu-Ming and Bai, Jia-Ju and Su, Zhendong},
  booktitle={32nd USENIX Security Symposium (USENIX Security 23)},
  pages={4949--4965},
  year={2023}
}

@inproceedings{symbolic,
  title={Symbolic Execution for Dynamic Kernel Analysis},
  author={Pitigalaarachchi, Pansilu},
  booktitle={Proceedings of the 2024 on ACM SIGSAC Conference on Computer and Communications Security},
  pages={5104--5106},
  year={2024}
}

@article{zhong2025testing,
  title={Testing Database Systems with Large Language Model Synthesized Fragments},
  author={Zhong, Suyang and Rigger, Manuel},
  journal={arXiv preprint arXiv:2505.02012},
  year={2025}
}

@inproceedings{blazytko2019grimoire,
  title={$\{$GRIMOIRE$\}$: Synthesizing structure while fuzzing},
  author={Blazytko, Tim and Aschermann, Cornelius and Schl{\"o}gel, Moritz and Abbasi, Ali and Schumilo, Sergej and W{\"o}rner, Simon and Holz, Thorsten},
  booktitle={28th USENIX Security Symposium (USENIX Security 19)},
  pages={1985--2002},
  year={2019}
}

@inproceedings{fu2022griffin,
  title={Griffin: Grammar-free dbms fuzzing},
  author={Fu, Jingzhou and Liang, Jie and Wu, Zhiyong and Wang, Mingzhe and Jiang, Yu},
  booktitle={Proceedings of the 37th IEEE/ACM International Conference on Automated Software Engineering},
  pages={1--12},
  year={2022}
}

@article{yang2025autoverus,
  title={Autoverus: Automated proof generation for rust code},
  author={Yang, Chenyuan and Li, Xuheng and Misu, Md Rakib Hossain and Yao, Jianan and Cui, Weidong and Gong, Yeyun and Hawblitzel, Chris and Lahiri, Shuvendu and Lorch, Jacob R and Lu, Shuai and others},
  journal={Proceedings of the ACM on Programming Languages},
  volume={9},
  number={OOPSLA2},
  pages={3454--3482},
  year={2025},
  publisher={ACM New York, NY, USA}
}

@inproceedings{xu2025ckgfuzzer,
  title={CKGFuzzer: LLM-Based Fuzz Driver Generation Enhanced By Code Knowledge Graph},
  author={Xu, Hanxiang and Ma, Wei and Zhou, Ting and Zhao, Yanjie and Chen, Kai and Hu, Qiang and Liu, Yang and Wang, Haoyu},
  booktitle={2025 IEEE/ACM 47th International Conference on Software Engineering: Companion Proceedings (ICSE-Companion)},
  pages={243--254},
  year={2025},
  organization={IEEE}
}

@article{zhang2025unlocking,
  title={Unlocking Low Frequency Syscalls in Kernel Fuzzing with Dependency-Based RAG},
  author={Zhang, Zhiyu and Li, Longxing and Liang, Ruigang and Chen, Kai},
  journal={Proceedings of the ACM on Software Engineering},
  volume={2},
  number={ISSTA},
  pages={848--870},
  year={2025},
  publisher={ACM New York, NY, USA}
}

@article{coverup,
  title={CoverUp: Effective High Coverage Test Generation for Python},
  author={Altmayer Pizzorno, Juan and Berger, Emery D},
  journal={Proceedings of the ACM on Software Engineering},
  volume={2},
  number={FSE},
  pages={2897--2919},
  year={2025},
  publisher={ACM New York, NY, USA}
}

@article{mutap,
  title={Effective test generation using pre-trained large language models and mutation testing},
  author={Dakhel, Arghavan Moradi and Nikanjam, Amin and Majdinasab, Vahid and Khomh, Foutse and Desmarais, Michel C},
  journal={Information and Software Technology},
  volume={171},
  pages={107468},
  year={2024},
  publisher={Elsevier}
}

@article{sqlancerplus,
  title={Scaling Automated Database System Testing},
  author={Zhong, Suyang and Rigger, Manuel},
  journal={arXiv preprint arXiv:2503.21424},
  year={2025}
}

@misc{sqlsmith2018,
  author = {Seltenreich, A. and Tang, B. and Mullender, S.},
  title = {SQLsmith: A random SQL query generator},
  year = {2018},
  howpublished = {\url{https://github.com/anse1/sqlsmith}}
}

@article{apollo,
  title={Apollo: Automatic detection and diagnosis of performance regressions in database systems},
  author={Jung, Jinho and Hu, Hong and Arulraj, Joy and Kim, Taesoo and Kang, Woonhak},
  journal={Proceedings of the VLDB Endowment},
  volume={13},
  number={1},
  pages={57--70},
  year={2019},
  publisher={VLDB Endowment}
}

@misc{sqlancer,
  author = {M. Rigger},
  title = {Sqlancer: Detecting Logic Bugs in DBMS},
  year = {2020},
  howpublished = {\url{https://github.com/sqlancer/sqlancer}}
}

@article{whitebox,
  title={Advanced white-box heuristics for search-based fuzzing of rest apis},
  author={Arcuri, Andrea and Zhang, Man and Galeotti, Juan},
  journal={ACM Transactions on Software Engineering and Methodology},
  volume={33},
  number={6},
  pages={1--36},
  year={2024},
  publisher={ACM New York, NY}
}

@article{advancing,
  title={Advancing code coverage: Incorporating program analysis with large language models},
  author={Yang, Chen and Chen, Junjie and Lin, Bin and Wang, Ziqi and Zhou, Jianyi},
  journal={ACM Transactions on Software Engineering and Methodology},
  year={2024},
  publisher={ACM New York, NY}
}

@article{shaohuali,
  title={Interleaving Large Language Models for Compiler Testing},
  author={Ni, Yunbo and Li, Shaohua},
  journal={Proceedings of the ACM on Programming Languages},
  volume={9},
  number={OOPSLA2},
  pages={815--841},
  year={2025},
  publisher={ACM New York, NY, USA}
}

@article{gai2025llama,
  title={LLAMA: Multi-Feedback Smart Contract Fuzzing Framework with LLM-Guided Seed Generation},
  author={Gai, Keke and Liang, Haochen and Yu, Jing and Zhu, Liehuang and Niyato, Dusit},
  journal={arXiv preprint arXiv:2507.12084},
  year={2025}
}

@inproceedings{xia2024fuzz4all,
  title={Fuzz4all: Universal fuzzing with large language models},
  author={Xia, Chunqiu Steven and Paltenghi, Matteo and Le Tian, Jia and Pradel, Michael and Zhang, Lingming},
  booktitle={Proceedings of the IEEE/ACM 46th International Conference on Software Engineering},
  pages={1--13},
  year={2024}
}

@article{luo2025enhancing,
  title={Enhancing protocol fuzzing via diverse seed corpus generation},
  author={Luo, Zhengxiong and Du, Qingpeng and Wang, Yujue and Roychoudhury, Abhik and Jiang, Yu},
  journal={IEEE Transactions on Software Engineering},
  year={2025},
  publisher={IEEE}
}

@inproceedings{cheng2025rug,
  title={RUG: Turbo LLM for Rust Unit Test Generation},
  author={Cheng, Xiang and Sang, Fan and Zhai, Yizhuo and Zhang, Xiaokuan and Kim, Taesoo},
  booktitle={2025 IEEE/ACM 47th International Conference on Software Engineering (ICSE)},
  pages={634--634},
  year={2025},
  organization={IEEE Computer Society}
}

@misc{doxygen,
  title        = {Doxygen},
  howpublished = {\url{https://doxygen.cn/index.html}},
  year         = {2026}
}

@misc{pin,
  title        = {Pintool},
  howpublished = {\url{https://www.intel.com/content/www/us/en/developer/articles/tool/pin-a-dynamic-binary-instrumentation-tool.html}},
  year         = {2026}
}

@misc{kes,
  title        = {KingbaseES},
  howpublished = {\url{https://www.kingbase.com.cn/}},
  year         = {2026}
}

@misc{neo4j,
  title        = {Neo4j},
  howpublished = {\url{https://neo4j.com/}},
  year         = {2026}
}

@article{gpt4,
  title={Gpt-4 technical report},
  author={Achiam, Josh and Adler, Steven and Agarwal, Sandhini and Ahmad, Lama and Akkaya, Ilge and Aleman, Florencia Leoni and Almeida, Diogo and Altenschmidt, Janko and Altman, Sam and Anadkat, Shyamal and others},
  journal={arXiv preprint arXiv:2303.08774},
  year={2023}
}

\end{document}